\documentclass[preprint,prb,showpacs,floatfix]{revtex4}  
\usepackage{graphicx}
\usepackage{amssymb}

\def\beeq{\begin{equation}}
\def\eneq{\end{equation}}
\def\beeqa{\begin{eqnarray}}
\def\eneqa{\end{eqnarray}}

\begin{document}

\title{Nucleation and crystallization process of silicon using
  Stillinger-Weber potential} 
 
 \author{Philippe Beaucage} 
 \author{Normand Mousseau } \email {Normand.Mousseau@umontreal.ca}
 \affiliation{D\'epartement de physique and Regroupement qu\'eb\'ecois
   sur les mat\'eriaux de pointe, Universit\'e de Montr\'eal, C.P.
   6128, succ. Centre-ville, Montr\'eal (Qu\'ebec) H3C 3J7, Canada}

\date{\today}

\begin{abstract}
  We study the homogeneous nucleation process in Stillinger-Weber
  silicon in the NVT ensemble.  A clear first-order transition from
  the liquid to crystal phase is observed thermodynamically with
  kinetic and structural evidence of the transformation. At 0.75
  $T_m$, the critical cluster size is about 175 atoms.  The lifetime
  distribution of clusters as a function of the maximum size their
  reach follows an inverse gaussian distribution as was predicted
  recently from the classical theory of nucleation (CNT). However,
  while there is a qualitative agreement with the CNT, the free energy
  curve obtained from the simulations differs significantly from the
  theoretical predictions, suggesting that the low-density liquid phase
  found recently could play a role in the nucleation process.
\end{abstract}


\maketitle
\section{Introduction}
\label{intro}

The classical nucleation theory (CNT) has been extensively tested in
systems with relatively simple two-body interactions such as colloids
or globular proteins ~\cite{Auer, Anderson, Yau, Wolde}.  These
molecules are large and move slowly, making it possible to follow the
crystallization process experimentally using various techniques of
microscopy.  Moreover, these systems can also be represented
accurately by theoretical models of hard- and soft-spheres, which can
crystallize on numerical time scales. It is therefore possible to
characterize fully the microscopic mechanisms responsible for
nucleation in terms of the CNT, which works particular well for these
systems. 

There has also been a number of studies going beyond the soft-sphere
models. In particular, there has been considerable work devoted to the
nucleation of Lennard-Jones models~\cite{Huitema, Baez, Swope,
  Honeycutt2, Honeycutt1}.  Very little work has been done, however,
on more complex materials such as oriented liquids --- water or
tetrahedral semiconductors, for example. Recently, Matsumoto {\it et al.}
~\cite{Matsumoto}, using considerable computing power, managed to
follow one occurrence of crystallization in a 300 ns run of a
512-molecule simulation of water in the canonical ensemble at 230 K.
Clearly more simulations are needed in water but also in simpler
oriented liquids such as silicon, which shows a similar
phases diagram around melting as  both liquids show a temperature of
density maximum and their density falls off by $\sim$ 10\% from the
disordered liquid to the tetrahedral crystalline structure.
As with water, there has been very few works studying nucleation in
this technologically important material~\cite{Nak}.

Depending on the cooling rate, previous numerical work has shown that
supercooled liquid silicon transforms in a glassy ~\cite{Luedtke,
  Broughton} or amorphous ~\cite{Angell, Luedtke2} state. Recently, it
was indicated that this transition takes place just below a
liquid-liquid transition~\cite{Sastry, Beaucage}: at zero pressure in
the Stillinger-Weber silicon, the low density liquid (LDL), which is
thermodynamically and structurally contiguous to the amorphous solid,
crystallizes rapidly (around 10 ns) at 1050 K~\cite{Sastry, Beaucage}
whereas the more common high-density liquid (HDL) does not at any
temperature on a simulation timescale. In order to circumvent the
difficulty to crystallize {\it l}-Si, Uttormark and colleagues
~\cite{Uttormark} embedded a spherical crystal seed containing 400-800
atoms in bulk liquid and analyzed the growth and dissolution of
clusters. They found that the critical size for a crystallite to grow
to macroscopic size was of 140 and 1400 atoms at 60\% and 85\% of the
melting temperature (T$_m$).  Working with a similar method, Bording
and Taft\o ~\cite{Bording} inserted a crystallite in an amorphous
matrix of 4096 germanium atoms and estimated the critical cluster
radius to be 2 nm (around 1500 atoms) at 60\% T$_m$.

In this paper, we show that liquid silicon can crystallize in the NVT
ensemble on timescale accessible by MD simulation without going
through the low-density liquid phase.  We also show that the
nucleation process, while qualitatively consistent with CNT, differs
quantitatively from it.  

The organization of the paper is as follow.  We show the behavior of
the thermodynamic, kinetic and structural properties during the phase
transition in section ~\ref{phase}. In section ~\ref{stability}, we
analyze the nucleation and crystallization process through the
evolution of the cluster that will eventually crystallize the whole
system in relation to CNT.  Then, in section ~\ref{free_e}, we compute
and compare the free energy of clusters between CNT and the simulation
data. Finally, we look at the lifetime of small clusters in the
supercooled liquid before nucleation takes place in section
~\ref{life}.

\section{Methods} \label{meth}

The molecular dynamical simulations (MD) for this work are performed
in the canonical (NVT) ensemble at the 0 K crystalline density, i.e.
2.32 g/cm$^3$, and in the isothermal-isobaric (NPT) ensemble at 0
pressure. All simulations are done at 1250 K (75\% T$_m$) in a cubic
box containing 10648 atoms, with periodic-boundary conditions.  This
size is sufficiently large to avoid catastrophic crystal growth due to
interactions between the images of the critical crystallite, which is
estimated to be around 200 atoms (see below).

We use the extended-system method of Andersen to control
pressure~\cite{Andersen, Haile, Brown} and Hoover's constraint method
for the temperature~\cite{Hoover, Evans, Allen}. Newton's equations of
motion are integrated with a fifth-order Gear predictor-corrector and
a time step $\Delta t = 1.15$ fs. Simulations are typically
equilibrated over 50 000 $\Delta t$ (58 ps) and data are accumulated
over $10^6 ~\Delta t$ (several ns). Atomic interactions are
represented by the Stillinger-Weber potential (SW), developed to
reproduce accurately the crystalline and liquid state of
Si~\cite{SW01}.

Starting with a liquid well equilibrated at 2900 K, we generate nine
independent trajectories in NVT conditions at 2.32 g/cm$^3$ and 75\%
of T$_m$, a degree of undercooling similar to that used for a wide
variety of materials both experimentally~\cite{Jackson} and
numerically.~\cite{Baez}  Of these nine trajectories, six crystallize
within 10 ns and are numbered 1 to 6;  the fastest, simulation \# 1,
crystallizes within 1.5 ns.

\begin{figure}[] 
\centering
\begin{tabular}{ccc}
\includegraphics[width=0.9in]{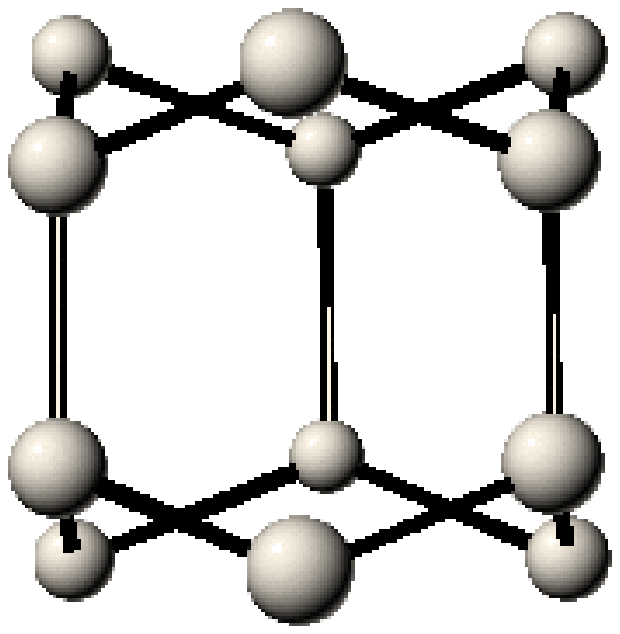} & 
\includegraphics[width=1.0in]{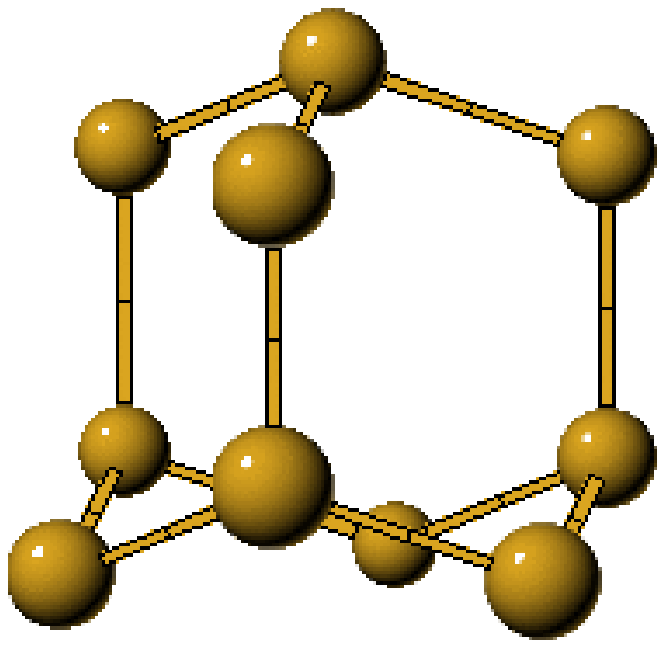} &
\includegraphics[width=1.0in]{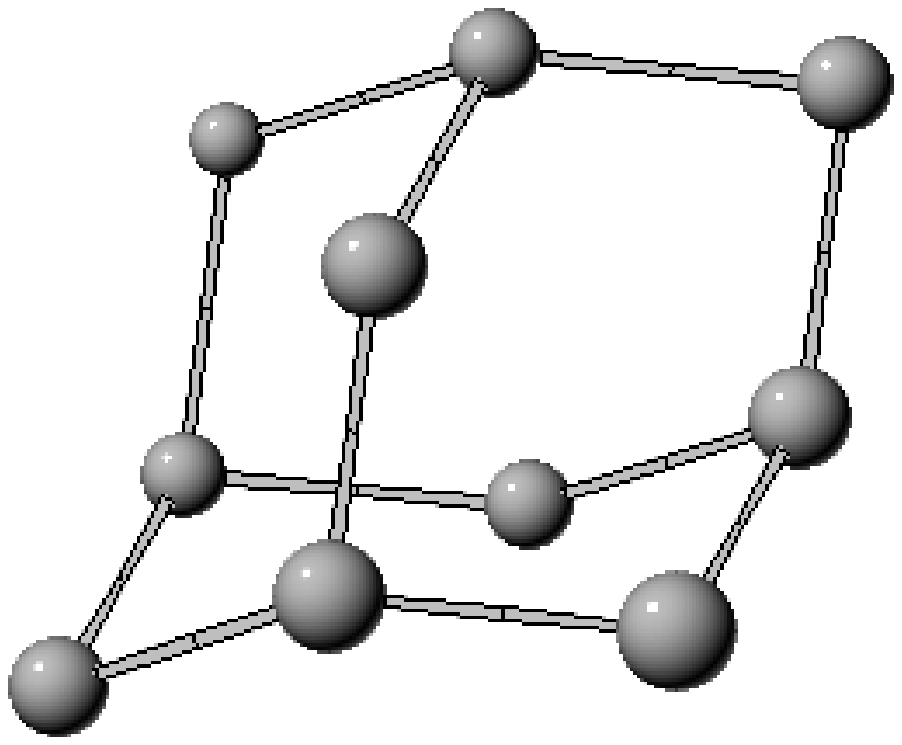} 
\end{tabular}
\caption{ (color online) The three basic building blocks associated with the
  crystalline order parameter. 
The wurtzite basic block (left) is a 12-atom cluster composed of two 
sixfold rings whereas the diamond basic block (middle) is a 10-atom 
cluster with four sixfold rings. The $\beta$-tin basic block (right) is 
equal to a diamond basic block where the tetrahedra 
are compressed in one direction and elongated along the two others axes.}
\label{basic_b}
\end{figure}

Following previous work on liquid Si~\cite{Nak, Beaucage}, we use as
order parameter the smallest three-dimensional closed-ring structures
that can be associated with a given crystalline lattice. These
clusters, shown in Fig.~\ref{basic_b}, are the smallest elementary
building blocks for wurtzite, diamond and $\beta$-tin structures and
are defined topologically: the wurtzite lattice is associated with a
12-atom cluster composed of two sixfold rings connected at three
points while the diamond and diamond can be described topologically by
a single 10-atom cluster with four sixfold rings back to back. To
establish the connectivity of these clusters, the first-neighbor
cut-off is set to 2.75 \AA, a value similar to that used in these
high-quality amorphous networks. This is somewhat shorter than the
typically nearest-neighbor distance used in liquids (which is about
3.0 \AA) as it focuses on local crystalline order.

These elementary clusters are present with a low density in the liquid
($\rho_{\rm crystal} \approx 5-10$ \% at.) as well as in high-quality
amorphous models prepared using the modified WWW bond-switching
algorithm ($\rho_{\rm crystal} \approx 1-5$ \% at.)~\cite{Barkema00}.
These blocks provide therefore a much more sensitive measure of
crystallinity than the structure factor or the RDF.

Our criteria are different from those used in a previous study of the
nucleation of crystallites implanted into a SW liquid by Uttormark,
Thompson and Clancy~\cite{Uttormark}. In this case, the description of
a crystallite nucleus is defined uniquely based on a mixture of
energetic, topological and geometric constraints. For an atom to be
part of a crystallite: (i) its three-body energy in SW potential of
fourfold or fivefold coordinated atoms (within a 3.35 \AA
~nearest-neighbor distance) must be lower than 0.4336 eV; (ii) it must
posses four nearest neighbors and at least three of them are also
fourfold coordinated; (iii) its angular bond angles meet this
criterion: $\sum_{i=1}^{6}(\cos \Theta_i + 1/3)^2 < 0.4$ (where
$\Theta_i$ is the angle between nearest neighbors of a fourfold
coordinated atom.)
The crystallites identified with this method are less compact than
those flagged with our topological order parameter. This is
particularly true for small crystallites (less than 20 atoms), which
tend to be open and stringy, like twisted polymers, with Uttormark's
criteria. The two methods converge, however, for larger clusters, near
and beyond the critical size, where a clear definition of surface is
less important.

\section{Results and Discussion}
\label{resul}

\subsection{Phase transition}
\label{phase}

Homogeneous nucleation is often difficult to obtain numerically,
especially in oriented solids such as Si and water which display a
crystalline structure far from that of the liquid phase. It took
months of computer time to simulate homogeneous nucleation in TIP3P
water.  Studies using SW Si failed to find traces of nucleation in a
5000-atom cell after a 1-ns simulation.~\cite{Uttormark}

In view of these results, and because classical nucleation theory
(CNT) ~\cite{Feder, Kashchiev} predicts that nucleation and
crystallization is obtained more rapidly for strong undercooling and
larger system size, we choose to simulate a larger cell, with more than
10$^4$ atoms, simulated over 10 ns at 0.75 $T_m$.

\begin{figure}[] 
\centering
\includegraphics[width=3.0in]{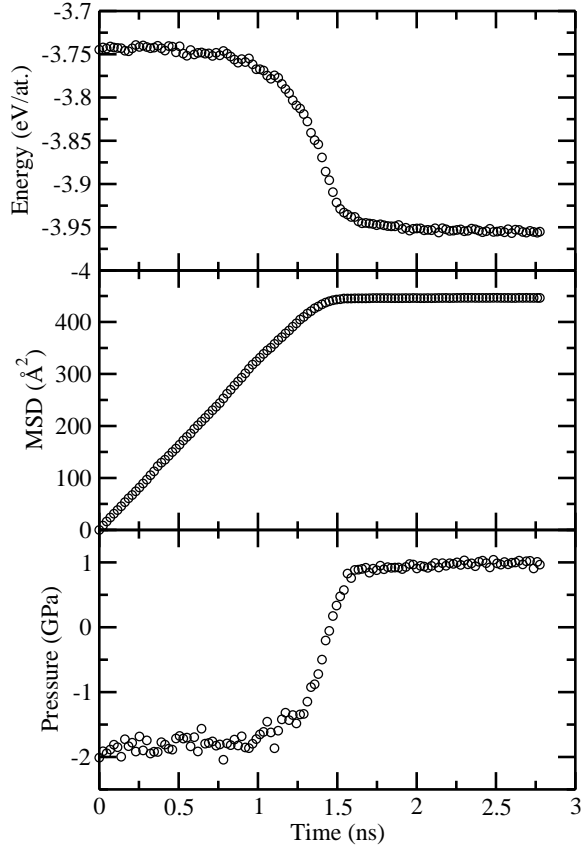} 

\caption{Evolution of the energy (top), mean square displacement
  (middle) and pressure (bottom) during the liquid-crystal phase
  transition of Si with NVT conditions at 1250 K and 2.32 g/cm$^3$.
  These results are for simulation \# 1, which crystallizes the fastest.
  While the other simulations take longer to crystallize, their
  evolution is similar.}
\label{thermo_NVT}
\end{figure}

As shown in Figure~\ref{thermo_NVT}, this is sufficient to observe
homogeneous nucleation, from the pure liquid phase, in the NVT
ensemble.  While the data presented in this figure are for simulation
\#1, a run that crystallizes particularly quickly, the overall
properties of the transition are identical to run \#2 to 6. The top
curve shows a brutal drop in the potential energy of the system, from
$-3.75$ to $-3.95$ eV/atom, indicating a clear thermodynamical
transition after 1.5 ns of simulation. The phase transition is also
visible by following the change in pressure (bottom panel). As the
density is maintained at the crystalline value, the pressure in the
liquid phase is negative; it changes sign at the liquid-crystal
transition since the crystal density at 1250 K and 0 GPa is slightly
lower than at 0 K, and since the new structure contains grain
boundaries. The liquid to solid phase transition is clearly seen in
the kinetics of the system (middle panel): in the supercooled liquid,
the diffusion is significant, with $D = 5.4 \cdot 10^{-6}$ cm$^2$/s;
it drops suddenly at the transition to become vanishingly small, a
clear indication of a liquid to solid transition.

Under the NVT conditions described in the introduction, the mean
pressure of the supercooled liquid is -1.9 GPa.  In a previous
work~\cite{Beaucage}, we studied the transition from high density
liquid (HDL) to low density liquid (LDL) in Stillinger-Weber Si and
showed that this transition does occur at around 1250 K and -2 GPa but
moves to lower temperatures as the pressure is increased. The current
simulations are therefore slightly above the HDL to LDL transition,
and we seem to observe a pure liquid-crystalline transition: the
liquid before the transition has a RDF and a diffusion constant
characteristic of the HDL and there is no trace of a LDL phase during
the crystallization process.

\begin{figure}[]
\centering
\includegraphics[width=2.5in, angle=-90]{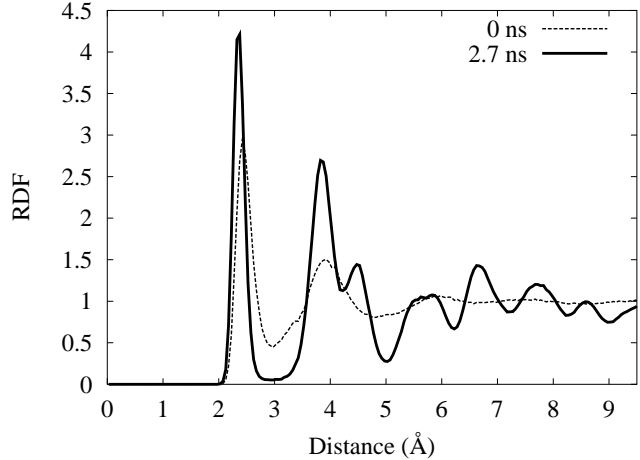} 
\caption{Radial distribution function before and after the liquid-crystal 
  phase transition of Si in NVT conditions at 1250 K and 2.32
  g/cm$^3$.  The RDF is characteristic of a crystalline state after
  the transition (2.7 ns) and of a liquid before the transition (0
  ns). These results are taken from Simulation \#1.}
\label{rdf_NVT}
\end{figure}

\begin{figure}[]
\centering
\includegraphics[width=2.5in]{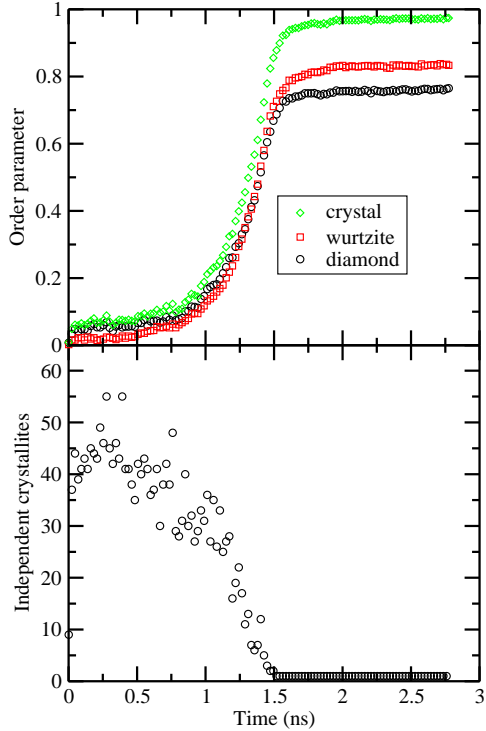} 
\caption{ (color online)
  Proportion of atoms in elementary blocks (top) and number of
  independent clusters (bottom) during the liquid-crystal phase
  transition of Si in NVT conditions at 1250 K and 2.32 g/cm$^3$. The
  proportion of atoms in diamond and/or wurtzite crystalline
  structures ($\diamondsuit$ crystal) increases rapidly reaching a
  value close to 1 after the transition.  These results are for
  Simulation \# 1.}
\label{struc_NVT}
\end{figure}

Changes in the structural properties of this model as the transition
occurs are shown in the next two figures.  At $t=0$ ns, the radial
distribution function (RDF) (see Fig.~\ref{rdf_NVT}) is typical of
that of a liquid, with little structure beyond the broad
second-neighbor peak.  The nature of the RDF is totally different
after the transition, with well-defined crystalline peaks up to 9 \AA\,
and beyond.  In the liquid phase, the system contains very few
crystalline building blocks and $\rho_{\rm crystal}$ fluctuates
between 5 and 10 \% of all the atoms (see Fig.~\ref{struc_NVT}, top
panel).  After the transition, more than 85 \% of atoms belongs to a 
diamond and/or wurtzite crystalline blocks, with a probability higher
for diamond structures except in trajectory \#1.
 
The co-existence of two crystalline structures is not surprising
since, with a cutoff of 3.77~\AA, the SW potential cannot
differenciate between the diamond and wurtzite structures at zero
temperature: these two structures start to differ only at their
third-neighbor shell, at 4.50 and 3.91 \AA, respectively.  It is
therefore only the thermal vibrations, bringing the third-neighbor
shell atoms inside the cut-off from time to time, that allow the
potential to distinguish between these two crystalline structures.
With long enough annealing, we expect the wurtzite structures to
disappear completely.

For its part, the liquid phase is characterized by a low density of
crystalline building blocks. Moreover, these crystallites tend to be
small, counting less than 20 atoms, on average. Before crystallization 
begins, the number of independent nuclei
oscillates between 40 and 50.  As crystallization occurs, however, the
largest nucleus grows rapidly, absorbing the smaller crystallites and
forming a single system-size cluster; the number of independent
crystallites decreases constantly during this process
(Fig.\ref{struc_NVT}, bottom panel).

\subsection{Characterization}

\subsubsection{Stability of crystallites}
\label{stability}

\begin{table}[]
\begin{center}
\caption{Characteristic times of the crystalline precursor that gives rise
  to crystallization of the supercooled liquid.} 
\begin{tabular}{|c|c|c|c|c|c|}
  \hline
  Simulations & \multicolumn{5}{c|}{Time (ns)} \\
  \cline{2-6}
   &  $t_0$ & $t_{200}$ & $t_{\rm nuc}$ & $t_{500}$ & $t_{\rm crys}$ \\ 
   \hline 
1 & 0.06 & 0.38 & 0.46 &  0.59 & 1.60 \\ 
2 & 0.64 & 0.75 & 0.84 &  1.02 & 2.75 \\
3 & 3.14 & 3.30 & 3.32 &  3.62 & 4.85 \\
4 & 3.30 & 3.67 & 4.20 &  4.68 & 6.00 \\
5 & 5.14 & 5.35 & 5.32 &  5.76 & 8.00 \\
6 & 7.79 & 7.98 & 7.98 &  8.27 & 9.75 \\
\hline
\end{tabular}
\label{carac1}
\end{center}
\end{table}

It is possible to characterize more finely the crystallization by
following the crystalline precursor as it takes over the simulation cell.
This is achieved by following the evolution of all crystallites by
steps of 1.15 ps. During this short time some crystallites appear,
other vanish, while the rest might evolve significantly; a set of
rules must therefore be established to identify uniquely and
reversibly each aggregate: (1) At least three atoms must remain
together over one time interval for a cluster to survive; a failing
test indicates that the aggregate has dissolved.  (2) When two or more
crystallites merge together, the one with the highest number of
surviving atoms is considered the progeny, the other one ceases to
exist. (3) If, on the other hand, a cluster splits into multiple
parts, the new aggregate containing the highest number of original
atoms becomes the progeny and the other clusters are considered
newborn.  Using this analysis, we can then follow the evolution of the
crystalline precursor by tracing back its ancestors.

\begin{figure}
\centering
\includegraphics[width=2.5in, angle=-90]{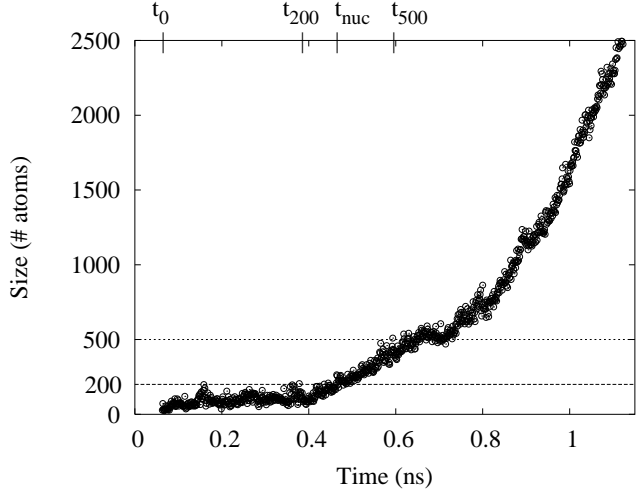} 
\caption{Evolution of the crystalline precursor during the liquid-crystal
  transition of Si in NVT conditions at 1250 K and 2.32
  g/cm$^3$. These results are taken from the first simulation (\#1) to
  crystallize.} 
\label{nuc01}
\end{figure}


In order to compare between the 6 runs that crystallize, we separate
the time evolution into four periods.  The instant of birth of the crystalline 
precursor is defined as $t_0$ (see Tab.~\ref{carac1}). From
this time, it may take several hundreds of picoseconds (about 200 to
900 ps) for this embryo to reach a critical size, at time  $t_{\rm nuc}$. The
nucleation time $t_{\rm nuc}$ is defined as the point
in time where the size of the aggregate starts growing
steadily, as seen in Fig.~\ref{nuc01}.  At this point, the system
leaves the incubation regime to enter the steady-stade of nucleation
and crystallization as such takes place.

CNT predicts that a cluster of over-critical size should grow
continuously whereas under-critical size crystallites tend to
dissolve, in both cases, to lower their free energy. Statistical
fluctuations can foil those predictions around the critical size,
however, and move from under-critical to over-critical size and
vice-versa.  This explains why we define $t_{\rm nuc}$ not as the
first time when the cluster reaches the critical size, but the first
time it reaches it for good. For example, while at $t_{\rm nuc}$
crystalline precursors are composed in average of 160 atoms, they have
often reached a size of 200 atoms or more before.  This fine
characterization of $t_{\rm nuc}$ is probably not needed, however.
Looking at Table~\ref{carac1}, $t_{\rm nuc}$ appears closely
correlated with $t_{200}$, the point in time where the crystallite
reaches a maximum size of 200 atoms for the first time.  The number of
clusters reaching a 200-atom size or more and then dissolving into
the liquid is extremely small. Thus, the critical cluster size 
should be around 175 atoms  for Si at 1250 K, in agreement with the
estimate of Uttormark {\it et al.}~\cite{Uttormark}.

>From $t_{\rm nuc}$, the crystallization {\it per se} proceeds rapidly
into a steady growth regime which lasts about 2 ns.  The
crystallization time, $t_{\rm crys}$, is defined as the moment when
the size of the largest cluster stops growing.


For all simulations, it is possible to trace back the critical cluster
to its appearance as a small aggregate of about 20 atoms, at $t_0$.
By selection, this cluster should live longer than most other
under-critical crystallites. As shown in Fig.~\ref{nuc01}, the size of
this cluster typically oscillates for a long time, aggregating and
loosing atoms until it reaches a critical size at $t_{nuc}$ and then 
starts growing for good. 

Surprisingly, while the cluster size oscillates, its composition
changes considerably. Throughout the incubation regime, the
crystalline precursor changes its composition significantly: very few
atoms of the original cluster remain part of it until the nucleation
phase starts.  In half the simulations, less than 50\% of the original
atoms are part of the cluster for 90\% of the time interval between
$t_0$ and $t_{\rm nuc}$ (see tab.~\ref{carac2}).  Even in the
steady-growth regime, starting at $t_{\rm nuc}$, the crystallite
continues to exchange atoms with the liquid. For most of the runs,
less than half the 160 or so atoms present at $t_{\rm nuc}$ remain in
the clusters for 90\% of the time in this interval until $t_{500}$; as
the growth takes place a significant fraction of the atoms move back
and forth between the crystallite and surrounding. These results are
in line with a previous study on growth and dissolution of implanted
LJ crystallites with a critical size similar to that of our
system~\cite{Baez} which shows that the probability of dissolution,
while decrease rapidly with cluster size above the critical size is
still non-negligible for clusters 50\% bigger than critical size.
Although we follow a cluster that will not dissolve totally, the
considerable atomic exchange is a reflection of this tendency.  While
the critical aggregate's composition changes rapidly, its position
remains almost fixed in space, its center of mass hardly moving except
by aggregation.  The crystalline precursor is therefore not a static
crystalline seed slowly growing throughout the nucleation process;
there is a constant exchange of matter with the surrounding liquid
even for post-critical sizes.

\begin{figure}
\begin{center}
\begin{tabular}{cc}
\includegraphics[width=3.5in]{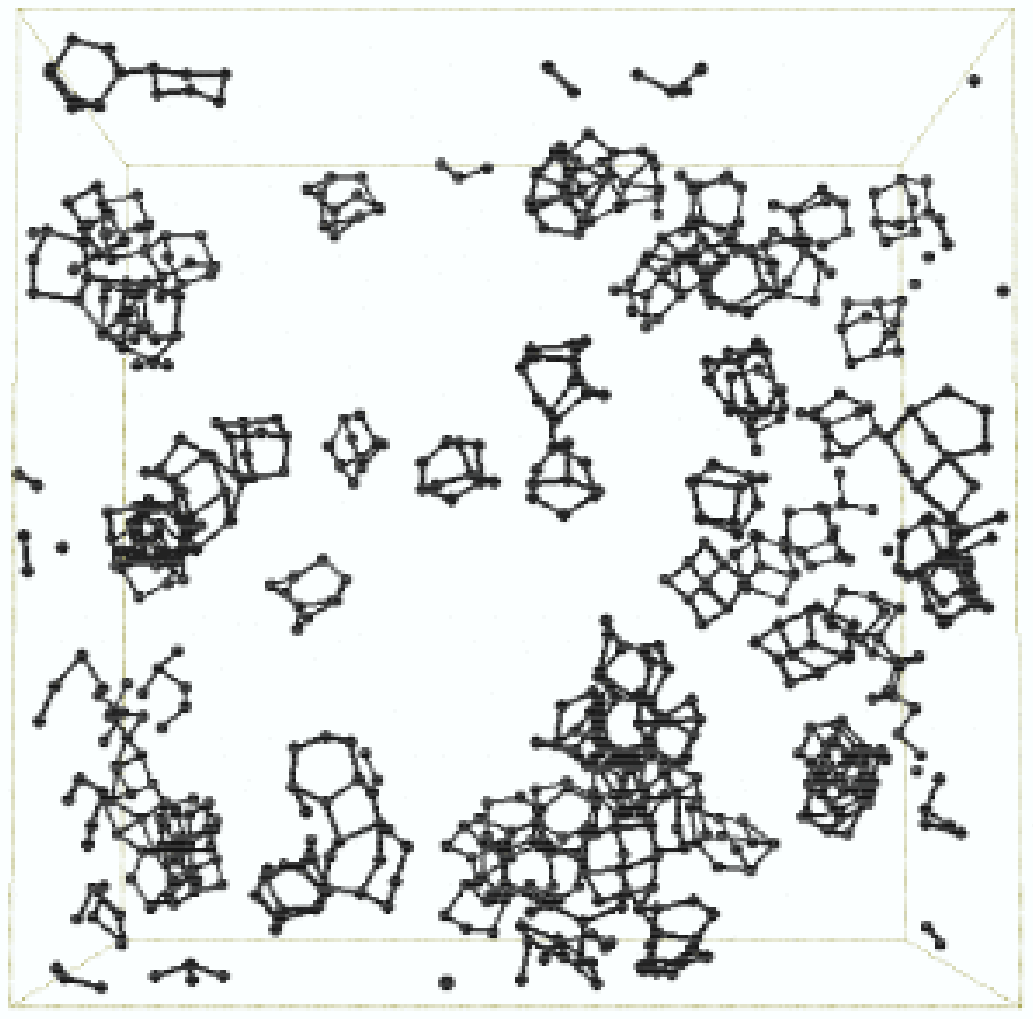} &
\includegraphics[width=3.5in]{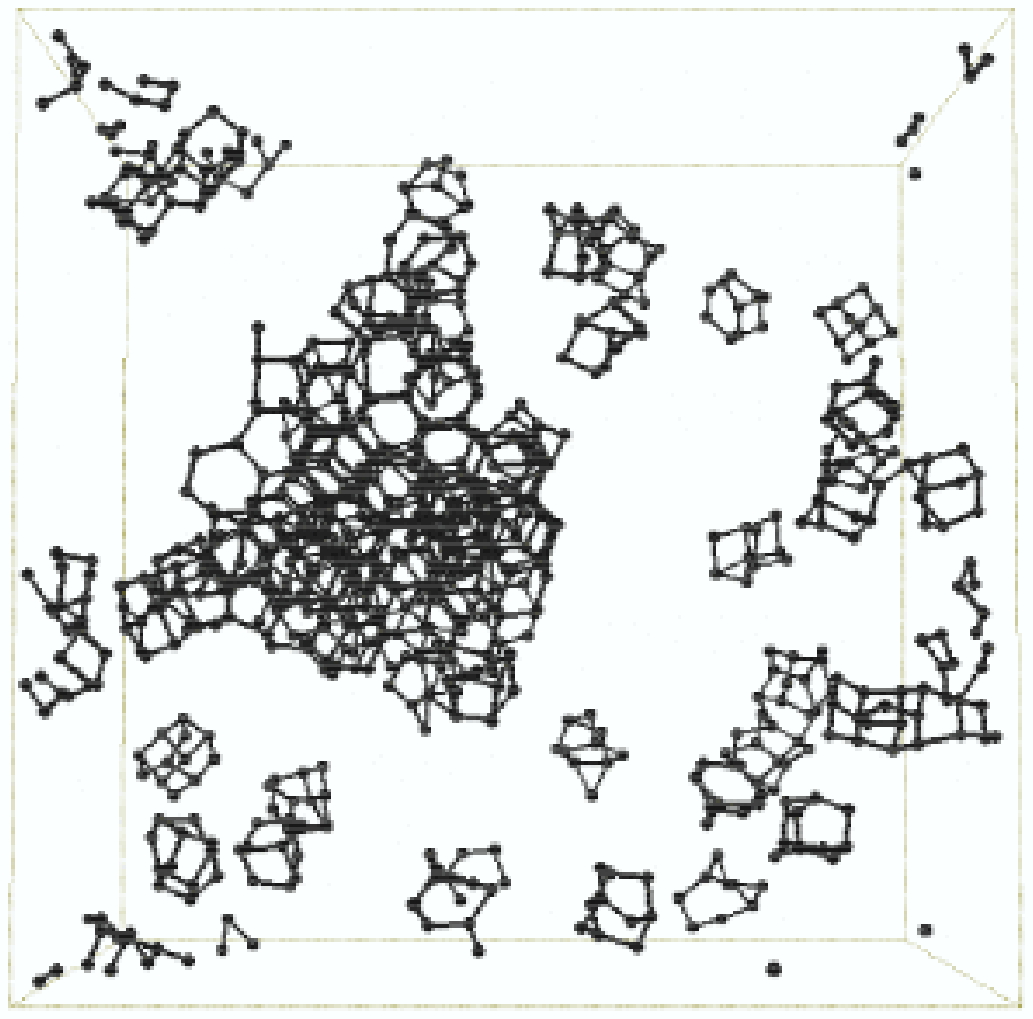}\\ 
\includegraphics[width=3.5in]{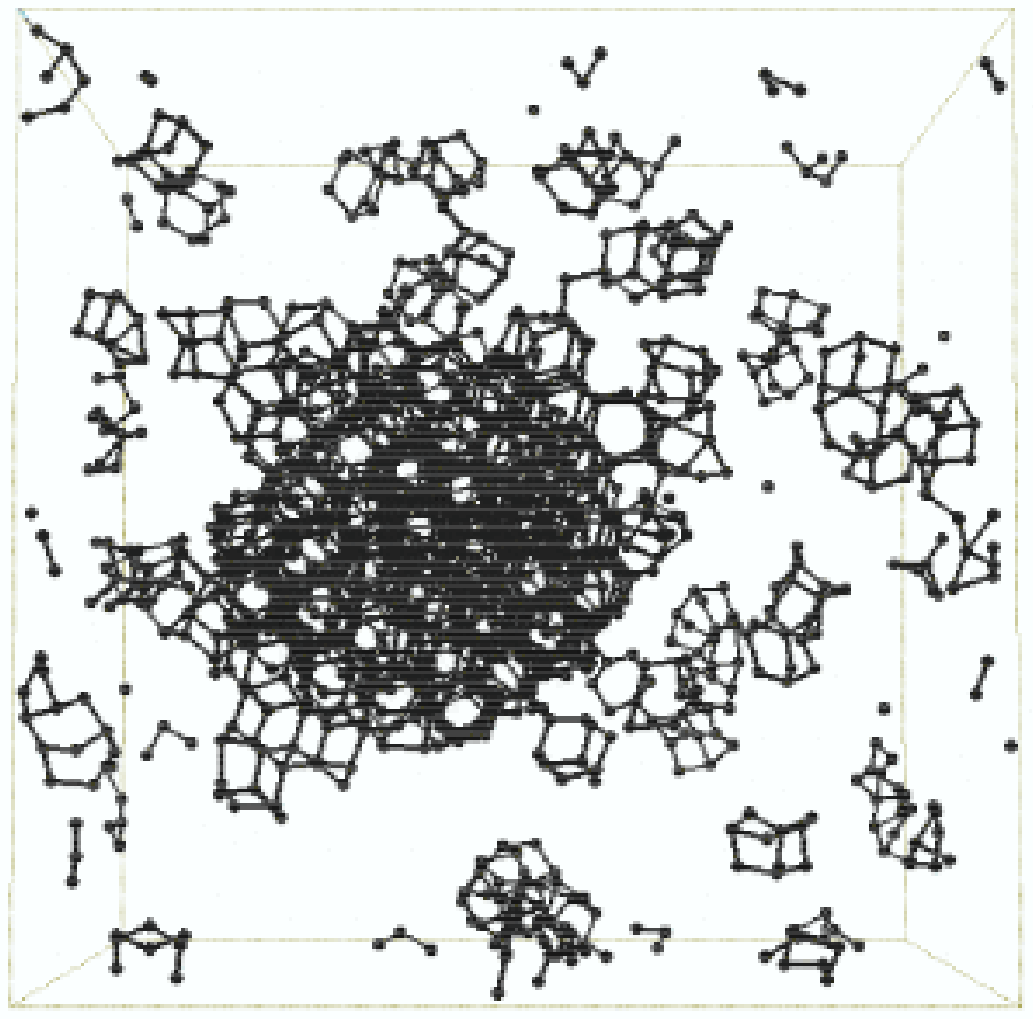}&
\includegraphics[width=3.5in]{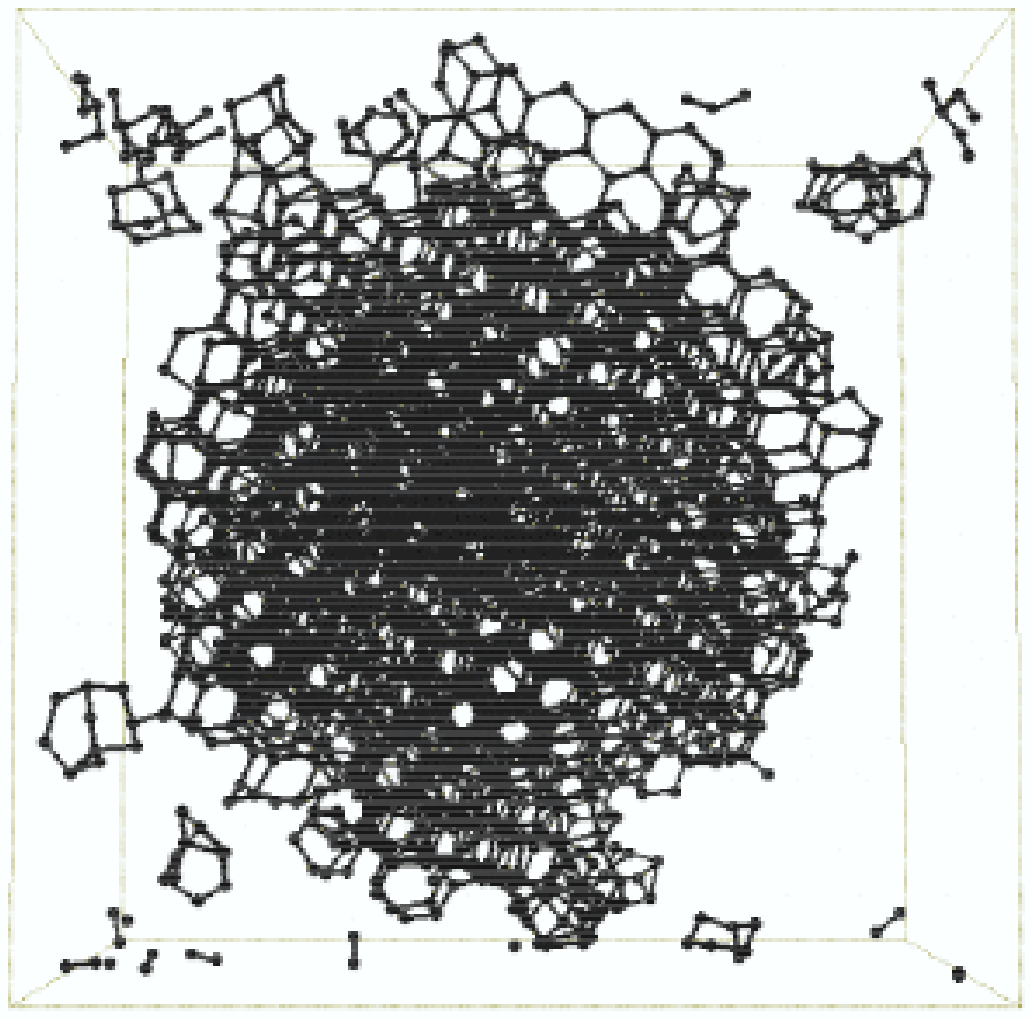} \\
\includegraphics[width=3.5in]{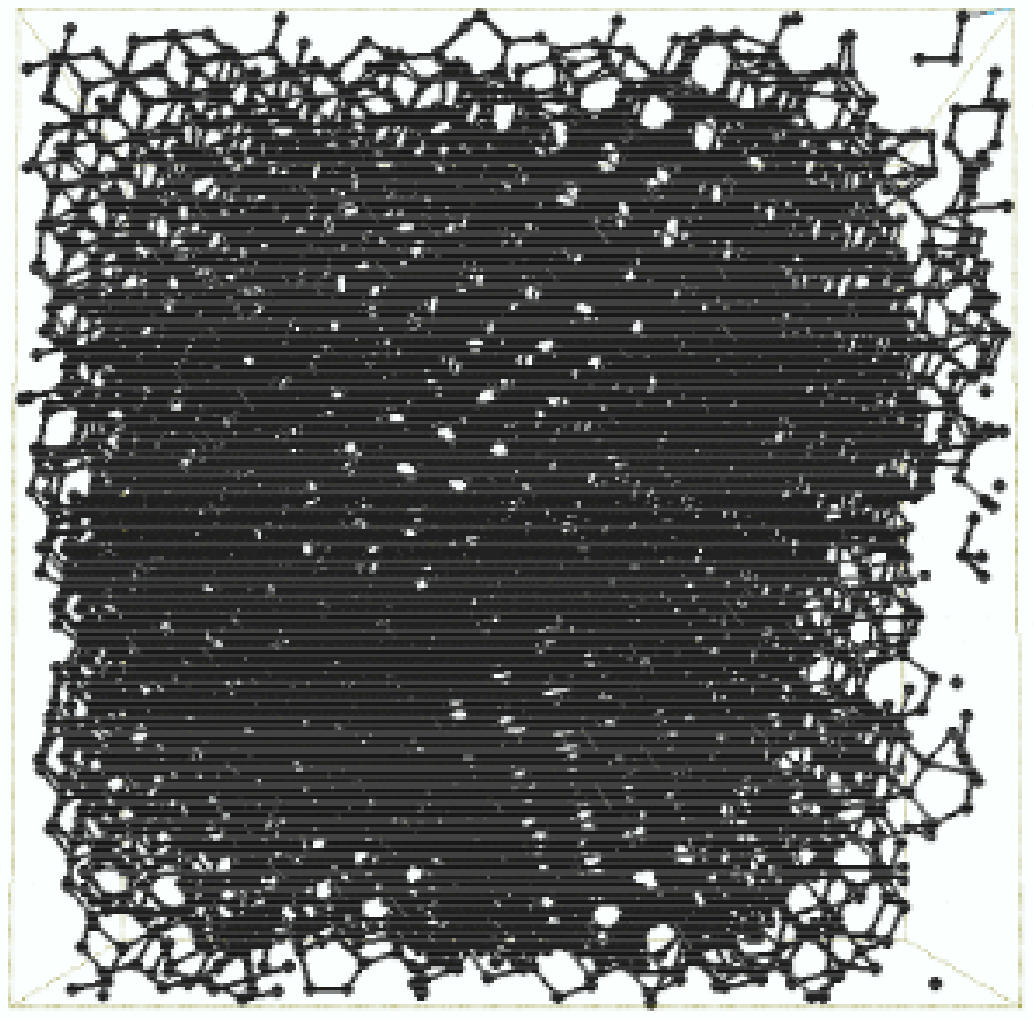} &
\includegraphics[width=3.5in]{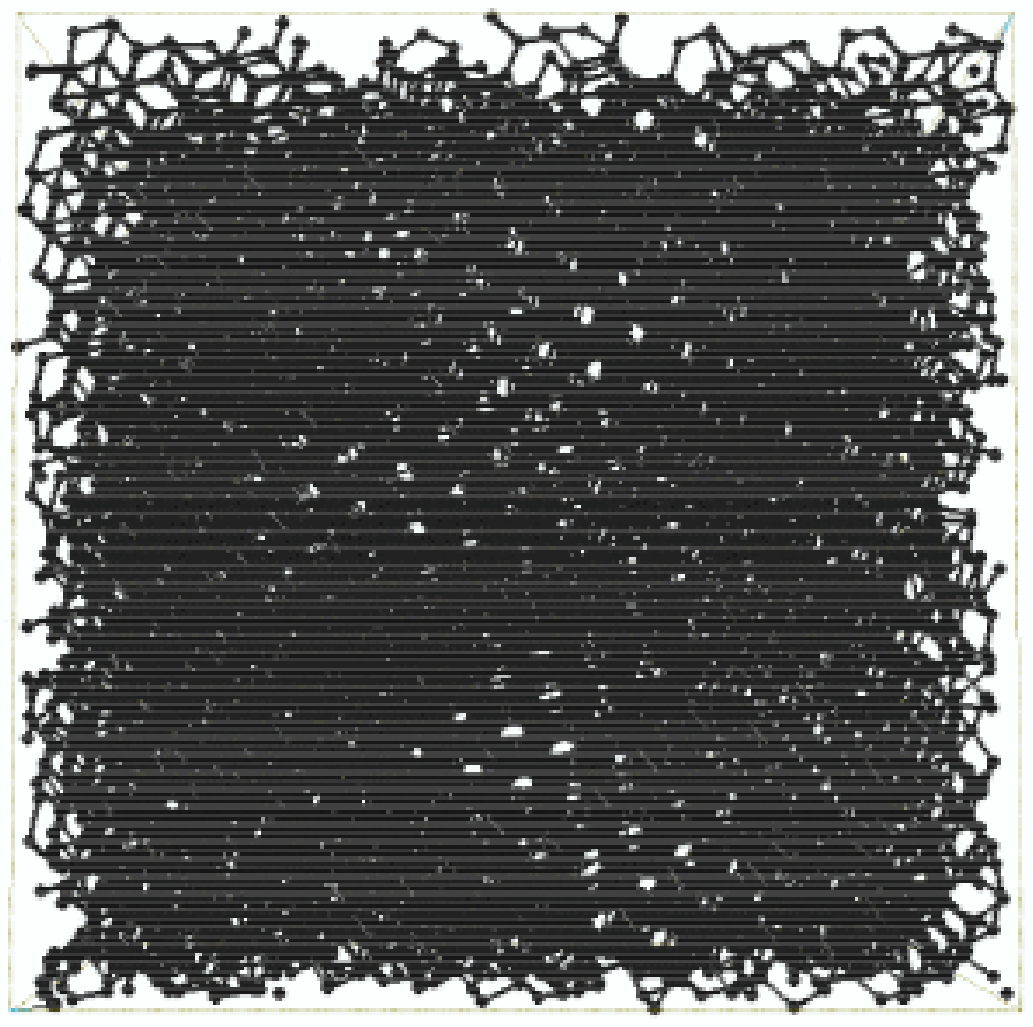} 
\end{tabular}
\end{center}
\caption{Evolution of nucleation and crystallization during the
liquid-crystal phase transition of SW Si at 1250 K and 2.32 g/cm$^3$. 
The configurations show atoms who belong only to crystalline structures
at 0, 0.58, 0.86, 1.15, 1.44 and 1.73 ns respectively for 
simulation \#1.}
\end{figure}

\begin{table}[]
\begin{center}
\caption{Proportion of atoms participating into the crystalline precursor permanently 
and 90 \% of the time during the incubation and steady-state regime 
of nucleation. Starting from atoms who belong originally to the 
crystallite at time $t_0$ until $t_{\rm nuc}$ in the incubation 
phase and from $t_{\rm nuc}$ until $t_{500}$ in steady-state. The 
interval between each configurations snapshot is 1.15 ps.} 
\begin{tabular}{|c|c|c|c|c|}
\hline 
Simulations & \multicolumn{4}{c|} {Persistence of atoms part of the crystalline precursor} \\
    \cline{2-5}         
& \multicolumn{2}{c|} {From $t_0$ to $t_{nuc}$} & \multicolumn{2}{c|}{From $t_{nuc}$ to $t_{500}$} \\
    \cline{2-5} 
& permanent & 90\% of the time & permanent &  90\% of the time\\
\hline
1 &  18\%    & 82\%  & 52\% & 81\%   \\
2 &   4\%    & 64\%  & 43\% & 77\%   \\
3 &  38\%    & 79\%  & 18\% & 57\%   \\
4 &   0\%    & 40\%  & 18\% & 51\%   \\
5 &   0\%    & 46\%  & 13\% & 40\%   \\
6 &   0\%    & 20\%  & 29\% & 49\%   \\
\hline
\end{tabular}
\label{carac2}
\end{center}
\end{table}

\subsubsection{Free energy}
\label{free_e}

It is formally straightforward to compare the simulations with the
predictions of CNT on the thermodynamics of crystal growth. The free
energy curve of crystallites can be obtained from the simulations by
plotting the equilibrium probability $P_{eq}(n)$ to find a crystallite
of size $n$ in the metastable liquid~\cite{Auer, Brendel}. 

We compute $P_ {eq}(n)$ in the supercooled liquid, accumulating data
until the largest cluster reaches 500 atoms, less than 5 \% of the
total number of atoms but over the critical size (see Tab.
\ref{carac1}), and over all runs.  This distribution is directly
connected to the free energy $\Delta F(n)$ associated with these
clusters:
\begin{eqnarray}
P_{eq}(n) &\propto& \exp\Big(\frac{-\Delta F(n)}{k_B T} \Big) \\
\frac{\Delta F(n)}{k_B T} &=& - \ln\Big(\frac{N(n)}{\sum_n N(n)}\Big) + C
\label{F_prob}
\end{eqnarray}
where $N(n)$ is the number of clusters of size $n$ present in the liquid, 
$k_B T$ is the Boltzmann constant times temperature and $C$, a constant.

\begin{figure}[] 
\centering
\includegraphics[width=2.5in, angle=-90]{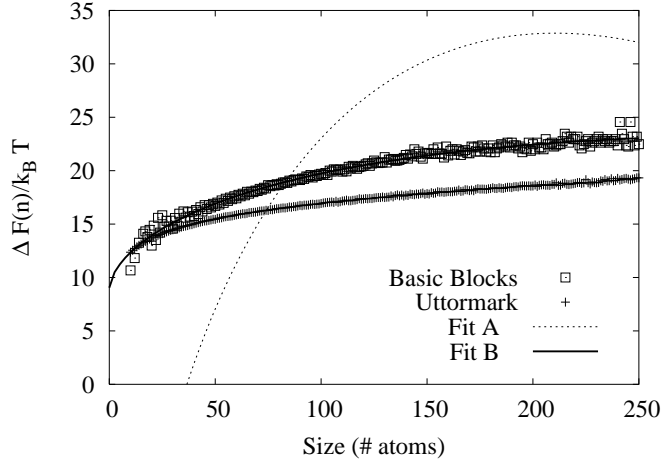} 
\caption{
  Free energy (divided by $k_B T$) of crystallites as a function of
  their size in the NVT ensemble. The simulation data are computed from
  the equilibrium probability of presence for clusters with the basic
  blocks analysis ($\Box$) or the criteria of Uttormark {\it et al.}  
  ~\cite{Uttormark} ($+$). 
  The CNT curve computed with the $\Delta F_{sl}$ value
  from thermodynamic integration is indicated as Fit A. A better fit
  is given by Fit B. Details are discussed in the text. }
\label{free_e01}
\end{figure}

The CNT offers another way to compute the free energy. In a simple
relation, the energy gain in the formation of a new phase is
balanced by the cost to produce an interface between the old and new
phases:
\begin{equation}
  \Delta F(n) =  \Delta F_{sl} \cdot n + \alpha \cdot n^{2/3} 
\label{tcn_01}
\end{equation}
where $\Delta F_{sl}= F_s- F_l$ is the Helmholtz free energy
difference between solid and liquid states in NVT conditions, $\alpha
= A \cdot \gamma$ with $\gamma$ the surface tension, $A = (36\pi /
\rho_s^2)^{1/3}$ for spherical crystallites and $\rho_s$ the density
of the solid phase.  While the Helmholtz free energy difference
$\Delta F_{sl}$ is relatively easy to obtain, the evaluation of the
surface tension is much trickier because small crystallites are far
from spherical and fluctuate considerably in shape for a given
size. Crystallites become mostly spherical only well beyond the
over-critical size.

Figure ~\ref{free_e01} compares the free energy for these two methods:
from the equilibrium probability (Eq.~\ref{F_prob}) and from CNT
predictions (Eq.~\ref{tcn_01}). Following standard practice, the
surface energy parameter $\alpha$ is fitted in order to obtain the
best agreement with the first method. The Helmholtz free energy
difference between the crystalline and liquid phases is computed as
follows.

The Gibbs free energy difference $\Delta G_{sl}$ between solid and
liquid states in NPT conditions at zero pressure is given by the
difference in chemical potential $\Delta \mu$ between the two phases.
This quantity was computed by Broughton and Li~\cite{Broughton} and
was found to be $-7.697 \cdot 10^{-2}$ eV/at. However, we need the
Helmholtz free energy difference $\Delta F_{sl}$ at fixed density,
which we can obtained by thermodynamical integration from the zero
pressure results. Starting with the relation for the internal pressure
$\Big(\frac{\partial F}{\partial V}\Big)_{N,T} = - P $, we use a
thermodynamic integration for each phase ($l$- and $s$-Si) :
\begin{equation}
  \Delta F = \int_{F_1}^{F_2} dF = - \int_{V_1}^{V_2} P(V) ~dV
\end{equation}

The free energy difference between our system at zero pressure and at
fixed density is computed by a Gaussian integration with five values:
\begin{equation}
  \int_{V_1}^{V_2} P(V) ~dV = \Big( \frac{V_2 - V_1}{2} \Big)
  \sum_{i=1}^{5} w_i P(V_i) 
\end{equation}
\begin{equation}
  V_i = \Big( \frac{V_2 - V_1}{2} \Big) x_i + \Big( \frac{V_2 + V_1}{2} \Big)
\end{equation}
where $x_i$ are the values for the Gaussian integration with their
relative weight $w_i$. The initial volume, at zero pressure, for the
liquid is $V_{1,l} = 18827.9$ ~\AA$^3$ (2.467 g/cm$^3$) and the solid,
$V_{1,s} = 20277.6$ ~\AA$^3$ (2.29 g/cm$^3$); the final volume is $V_2
= 20023.4$ ~\AA$^3$. Each point in the integral is simulated in NVT
conditions at 1250 K for the liquid and solid. We equilibrate our
1000-atoms system for 58 ps and then compute the mean pressure during
345 ps of simulation time.

After integrating, we find a free energy difference per
atom between the fixed density ($\rho$ = 2.32 g/cm$^3$) and zero
pressure system ~\cite{GPa_2_eV}, for liquid and solid state :
\begin{eqnarray}
\Delta F_s = 9.588 \cdot 10^{-4} ~eV/at. \\
\Delta F_l = 5.573 \cdot 10^{-3} ~eV/at.
\end{eqnarray}
This gives a free energy difference between the liquid and solid phase
at 2.32 g/cm$^3$ and 1250 K of
\begin{eqnarray}
\Delta F_{sl} &=& \Delta G_{sl} (P=0) + \Delta F_s - \Delta F_l \\
\Delta F_{sl} &=& -8.158 \cdot 10^{-2} ~eV/at. 
\label{deltaF}
\end{eqnarray}
The constant-volume correction is therefore only 6\% of the zero
pressure result of Broughton and Li.

As can be seen in Fig.~\ref{free_e01}, however, the CNT curve does not
match the free energy data coming from $P_{eq}(n)$ in simulations. In
order to find a better fit, the free energy difference $\Delta F_{sl}$
between the solid and liquid state should be nine times lower than the
value computed with the thermodynamic integration.

We can verify the impact due to the choice of the order parameter on the
free energy curve by re-analyzing the data using the criteria of
Uttormark {\it et al.}  ~\cite{Uttormark}.  The resulting curve is
also plotted in Fig.~\ref{free_e01} and shows an even flatter curve,
away from CNT results. We also repeated the simulation at 1250 K in
the NPT ensemble at zero pressure and over 10 ns. In this 
situation, the trajectories do not crystallize --- the largest crystallite
reaches about 100 atoms, well below the estimated critical size.
The free energy distribution obtained from the cluster size distribution,
while more curved than that for the NVT conditions, is still far from
the CNT predictions ($\Delta G_{sl}$ is about 5 times too low).

\begin{figure}[] 
\centering
\includegraphics[width=2.5in, angle=-90]{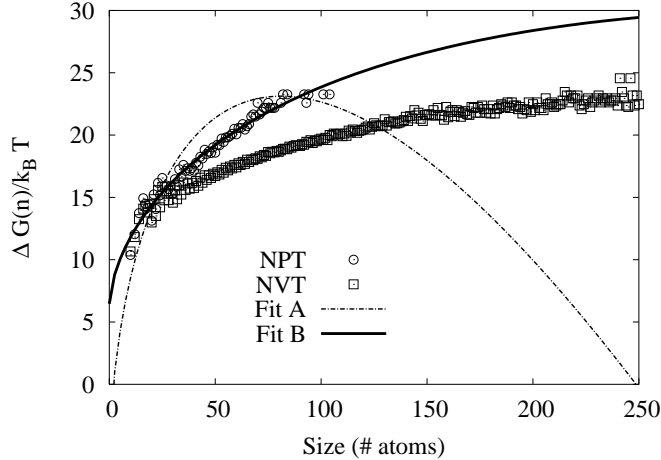} 
\caption{
  Free energy (divided by $k_B T$) of crystallites as a function of
  their size in NPT conditions (NPT) compared to NVT conditions (NVT). 
  The simulation data are computed from
  the equilibrium probability of presence for clusters with the basic
  blocks analysis. The CNT curve computed with the
  $\Delta G_{sl}$ of Broughton et Li (Fit A) is closer to the free
  energy data originating from $P_{eq}(n)$. However, a better fit (Fit
  B) requires a value five times lower.
  }
\label{free_e02}
\end{figure}

The discrepancy between the two approaches clearly indicates that the
classical nucleation theory does not fully capture the nucleation
process in SW Si. We identify two possible sources of discrepancy.
(1) As was demonstrated by Sastry and Angell
recently~\cite{Sastry,Beaucage}, SW Si undergoes a high-density to
low-density liquid-liquid phase transition. The low density phase
could be stabilized at higher temperature by the presence of a
crystallite. In this case, it would be necessary to take into account two
interfaces instead of one in the CNT equations. (2) The CNT fails
because the critical nucleus is too small breaking the approximation
of spherical crystallites. At this moment, we could not verify or
disprove either of these possibilities.

\subsubsection{Lifetime of crystallites}
\label{life}

Beyond the free energy curve, we also analyze the dynamics of the
crystallites present in the supercooled liquid.

The lifetime probability of crystallites can be derived by following
the kinetic approach of Zeldovich ~\cite{Feder}. This approach
predicts that the evolution of the clusters can be described by a
diffusion equation of the form:
\begin{equation}
 \frac{\partial c(n,t)}{\partial t} = \frac{\partial}{\partial n}
 \bigg\lbrace  D(n) \bigg[ \frac{\partial
 c(n,t)}{\partial n} + \frac{\partial \frac{\Delta G(n)}{k_B
 T}}{\partial n} \cdot c(n,t) \bigg]  \bigg\rbrace
\label{zeldo2} 
\end{equation}
where $c(n,t)$ is the concentration of crystallites of size $n$ at
time $t$, $D(n)$ the diffusion and $k_B ~T$ the boltzmann factor times
the temperature.

Van Kampen ~\cite{Kampen} resolved the differential equation for small
times by assuming the diffusion to be constant.  Further approximating
the potential as linear with respect to the cluster's size, van
Beijeren~\cite{Beijeren} succeeded in finding a solution for longer
times.  This latter equation, which gives the distribution function
for first arrival at size $n_f$, for crystallites starting from size
$n_0$ is a well-known results~\cite{Grimmett} that confirms van Kampen
short term behavior and contains an additional friction term
$e^{-\nu_0 t}$ which becomes important for longer times:
\begin{center}
\begin{equation}
P(n,n_f,t) = \frac{n_f-n_0}{\sqrt{4 \pi D t^3}} e^{- \frac{(n_f-n_0)^2}{4
D t}} e^{- \frac{(\Delta G(n_f) - \Delta G(n_0))}{2 k_B T}} e^{-\nu_0 t}
\label{Beij} 
\end{equation}
where
\begin{equation}
\nu_0 =  D \cdot \Big(\frac{ (\Delta G(n_f) - \Delta G(n_0))}{2 k_B T} \Big)^2. \nonumber
\end{equation}
\end{center}
This probability distribution is formally known as an inverse Gaussian
distribution (or inverse normal, Wald). It was first derived
independently by Schr\"odinger~\cite{Schro} and
Smoluchowski~\cite{Smolu} to describe Brownian motion in systems with
a drift velocity. Hence, the development of a crystallite can be
represented as a random walk in a field of force $\Delta F$ through
different size classes where small clusters have a strong tendency to
dissolve into the liquid (a drift to $n_f \rightarrow 0 $) and
super-critical nano-crystals tend to growth to macroscopic size ( $n_f
\rightarrow \infty $).  Since we do not have all the information on
the free energy of crystallites $\Delta G(n)$ (see
section~\ref{free_e}) and the diffusion constant, it is not possible
to use directly Eq.~\ref{Beij} to compare the lifetime behavior of
clusters in the supercooled liquid during nucleation. However, we can
circumvent the difficulty by writing the inverse Gaussian distribution
under a parametric form where $A$ represents the mean and $A/B^3$, the
variance:
\begin{center}
\begin{equation}
P(t) = \frac{B}{\sqrt{2 \pi t^3}} \exp \bigg( - \frac{B}{2t}
\Big(\frac{t-A}{A} \Big)^2 \bigg) 
\label{igd}
\end{equation}
\begin{eqnarray}
A &=& \frac{-(L-n_0)~k_B T}{D~(\Delta G(L) - \Delta G(n_0))} \\
\frac{A^3}{B}  &=& \frac{-2(L-n_0)}{D^2} \Big(\frac{k_B T}{\Delta G(L) - \Delta G(n_0)} \Big)^3
\end{eqnarray} 
\end{center}

\begin{figure}[] 
\centering
\includegraphics[width=2.5in,angle=-90]{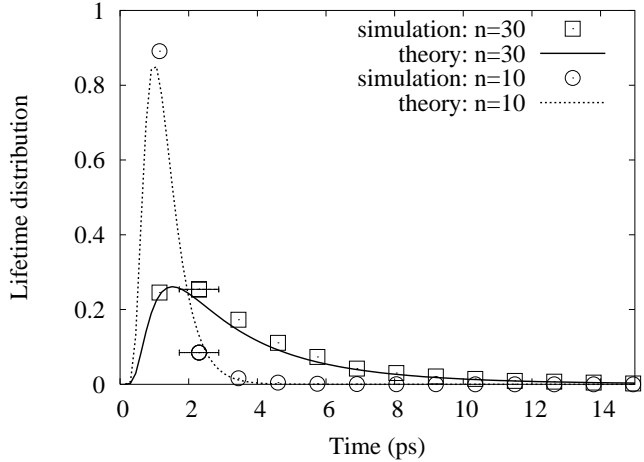} 
\caption{Lifetime distribution of clusters reaching a maximum size of 10 and 30
atoms. Comparison between the inverse gaussian distribution (theory) and the simulations data (simulation) with an uncertainty of $\pm$ 0.58 ps.}
\label{tv01}
\end{figure}

We compute the mean lifetime and variance for crystallites reaching
the same maximum size in order to determine the theoretical
distribution and compare with the lifetime probabilities from
numerical simulations. Because large clusters are not encountered
frequently, the amount of data collected over all MD simulations
remains small for the lilfetime of cluster near the critical size.  In
Fig.~\ref{tv01}, the lifetime distributions determined by the
Eq.~\ref{igd} and the simulations data are in good agreement for small
crystallites ensuring that cluster nucleation is well described by the
inverse Gaussian distribution. The mean lifetimes for crystallites
reaching an under-critical size of 10 or 30 atoms is 1.32$\pm$ 0.6 and
3.79 $\pm $ 0.6 ps respectively, with a variance of 1.73 and 10.01,
although some rare clusters last until 30 and 50 ps (not shown). As
would be expected, the mean lifetime increase with the size.

Although some approximations have been made to obtain the lifetime
probabilities of clusters by the inverse gaussian distribution and
from the simulations data, the results are conclusive for crystallites
reaching relatively small size. Since small clusters developed
themselves in a confined range of size, we believe that the free
energy difference can be approximated by a linear relation to the
crystallite size and the diffusion kept constant.

\section{Conclusions}
\label{concl}

There has been a lot of interest recently regarding the nature of the
liquid-solid transition in oriented liquid such as water and
tetrahedral semiconductors. In many systems, it appears that there
exists a high-density to low-density liquid transition often leading
to a glassy or amorphous phase~\cite{Sciortino,Sastry,Beaucage}. Here,
we reported results on a study of nucleation in liquid Si above the
HDL to LDL transition. 

We find that homogeneous nucleation takes place on a time scale of
about 10 ns in a large enough system at constant volume.  Using a
topological order parameter, it is possible to follow the evolution of
the crystallites through the crystallization process. Based on this
analysis, we estimate the critical size to be around 175 atoms, within
the limits of previous estimation of Uttormark {\it et al.}.
Surprisingly, the critical cluster, the one that will eventually
crystallize the whole system, can survive at under-critical size for a
long time (up to 900 ps or more) before it starts to grow steadily.
Although the cluster's center of mass does not move significantly,
there is a fluctuation in the composition of the cluster, as atoms
move from the liquid to the crystallite and vice-versa, even once the
crystallite has reached an over-critical size.

A comparison of the simulation results with the classical nucleation
theory indicates that the general behavior of the nucleation process
is in agreement with CNT. For example, we find that the lifetime
distribution of clusters reaching a specific maximum size follows the
inverse gaussian distribution predicted recently~\cite{Beijeren},
supporting the description of the cluster growth as a random walk in
the presence of a force field associated with the free energy.
However, the details of the nucleation free energy differ
significantly from the theoretical predictions. While the specific
origin of this discrepancy remains open, we suggest that it could be
caused by the presence of a low-density liquid at the interface
between the crystal and the normal liquid or by the small size of the
critical nucleus. More studies are required to fully address this
problem.

\section{Acknowledgments}

The authors want to thank G.T. Barkema, H. van Beijeren and K. Brendel
for helpful discussions.  This work is funded in part by NSERC, NATEQ
and the Canada Research Chair Program.  NM is a Cottrell Scholar of
the Research Corporation.  Most of the simulations were run on the
computers of the R\'eseau Qu\'eb\'ecois de Calcul de Haute Performance
(RQCHP) whose support is gratefully acknowledged.


\end{document}